\documentclass[a4paper]{PoS}
\usepackage{caption}
\usepackage{subcaption}
\usepackage{url}
\usepackage{lineno}
\newcommand{\doublefig}[9]{
	\begin{figure*}[t!]
		\begin{subfigure}[t]{0.49\textwidth}
			\fbox{\includegraphics[width=\textwidth,height=#1cm]{#2}}
			\caption{{\small #3.}}
			\label{#4}
		\end{subfigure}
		\begin{subfigure}[t]{0.49\textwidth}
			\fbox{\includegraphics[width=\textwidth,height=#1cm]{#5}}
			\caption{{\small #6.}}
			\label{#7}
		\end{subfigure}
		\caption{{\small #8.}}
		\label{#9}
	\end{figure*}
}
\newcommand{\onefig}[5]{
	\begin{figure}[t!]
		\centering
		\fbox{\includegraphics[width=#1cm,height=#2cm]{#3}}
		\caption{{\small #4.}}
		\label{#5}
	\end{figure}
}
\newcommand{\esqtab}[4]{
	\begin{table}
		\begin{center}
			\begin{tabular}{#3}
				#4
			\end{tabular}
			\caption{{\small #1.}}
			\label{#2}
		\end{center}
	\end{table}
}
\title{New method for Gamma/Hadron separation in HAWC using neural networks}
\ShortTitle{Gamma/Hadron separation in HAWC using NN}
\author{\speaker{T. Capistr\'an}$^a$, I. Torres$^a$, L. Altamirano$^a$ and for the HAWC Collaboration$^b$ \\
        \llap{$^a$}Instituto Nacional de Astrof\'isica, \'Optica y Electr\'onica, Luis Enrique Erro 1, Tonantzintla, Puebla 72840, M\'exico \\
        \llap{$^b$}For a complete author list, see \href{http://www.hawc-observatory.org/collaboration/icrc2015.php}{www.hawc-observatory.org/collaboration/icrc2015.php}.\\
        Email: \email{tcapistran@inaoep.mx}, \email{ibrahim@inaoep.mx}, \email{robles@inaoep.mx}}

\abstract{The High Altitude Water Cherenkov (HAWC) gamma-ray observatory is located at an altitude of 4100 meters in Sierra Negra, Puebla, Mexico. HAWC is an air shower array of 300 water Cherenkov detectors (WCD's), each with 4 photomultiplier tubes (PMTs). Because the observatory is sensitive to air showers produced by cosmic rays and gamma rays, one of the main tasks in the analysis of gamma-ray sources is gamma/hadron separation for the suppression of the cosmic-ray background. Currently, HAWC uses a method called compactness for the separation. This method divides the data into 10 bins that depend on the number of PMTs in each event, and each bin has its own value cut. 
In this work we present a new method which depends continuously on the number of PMTs in the event instead of binning, and therefore uses a single cut for gamma/hadron separation. The method uses a Feedforward Multilayer Perceptron net (MLP) fed with five characteristics of the air shower to create a single output value.
We used simulated cosmic-ray and gamma-ray events to find the optimal cut and then applied the technique to data from the Crab Nebula.  This new method is tuned on MC and predicts better gamma/hadron separation than the existing one. Preliminary tests on the Crab data are consistent with such an improvement, but in future work it needs to be compared with the full implementation of compactness with selection criteria tuned for each of the data bins.}

\FullConference{The 34th International Cosmic Ray Conference,\\
		30 July- 6 August, 2015\\
		The Hague, The Netherlands}
\begin{document}
	\section{Introduction}
		\paragraph*{} The High Altitude Water Cherenkov (HAWC) gamma-ray observatory is composed of 300 water Cherenkov detector (WCD). On the bottom of each WCD there are 4 photomultiplier tubes (PMTs) that detect the Cherenkov light. This light is produced by secondary particles in air shower generated by the interaction between atmosphere and primary particle (as for example gammas rays, protons, among other particles). The rate of cosmic rays (CR) is bigger than the gamma rays (GR) so it is critical to find a technique to remove the CR without losing the signals of GR.
		\paragraph*{}Currently, HAWC has a method called compactness for distinguishing those primary particles. For doing this, the data is divided into 10 bins (see Table~\ref{Tab:hawc300}) depending on $nHit$, that is the number of PMTs that have a signal in the event. The compactness depends upon the charge distribution deposited by the secondary particles of the shower on PMTs of the array. In this work, a new method is presented, using a Neural Network (NN) for the gamma/hadron separation without dividing the data into bins. Five characteristics are computed for feeding a NN that computes a value ($\theta_{NN}$) to distinguish between CR and GR. Another method in development can be found in~\cite{Zigicrc2015}.
	\section{Training stage}	
		\paragraph*{}The NN used in this work is a Feedforward Multilayer Perceptron~\cite{misc:pagrootmlp}. For a correct evaluation, the NN must pass two stages, training and testing. In the training stage the aim is to minimize the classification error. First, the values of characteristic input are calculated and a training MC data set is selected. The architecture is defined as 5-5-5-1 (Figure~\ref{Fig:estnn}), the first layer has 5 neurons because the NN need 5 characteristics as input\footnote{Each input is normalized with respect to maximum value of each feature.}, one neuron in the last layer because the network needs to recognize only two types of particle. Different architectures of NN were tested but the learning curves were similar. In the use of NN the recommended number of total layers should be $N-1$ where $N$ is the number of input variables \cite{book:NNchris}, in our case $N=5$ so the simple structure (5-5-5-1) was chosen to save computing time. The learning method used was stochastic minimization and took 500 epochs for a asymptotic behavior in the error of the output.
%\protect\cite{pacoicrc2015}
		\esqtab{nHit range and gamma/hadron cut in each bin for HAWC-300, $\theta_c$ is the compactness cut value}{Tab:hawc300}{|c|c|c|c|}{
			\hline
			bin &nHit min&nHit max&$\theta _c$\\
            \hline
			-1 & 30 & 54 & - \\
			\hline
			0 &  55 &   87 &  4.6\\
			\hline
			1 &  88 &  138 &  6.3\\
			\hline
			2 & 139 &  216 &  9.8\\
			\hline
			3 & 217 &  323 & 12.7\\
			\hline
			4 & 324 &  457 & 17.6\\
			\hline
			5 & 458 &  606 & 19.5\\
			\hline
			6 & 607 &  754 & 18.5\\
			\hline
			7 & 755 &  889 & 17.1\\
			\hline
			8 & 890 & 1000 & 15.0\\
			\hline
			9 & 1001 & 1200 & 12.4\\
			\hline
		}
		\doublefig{8}{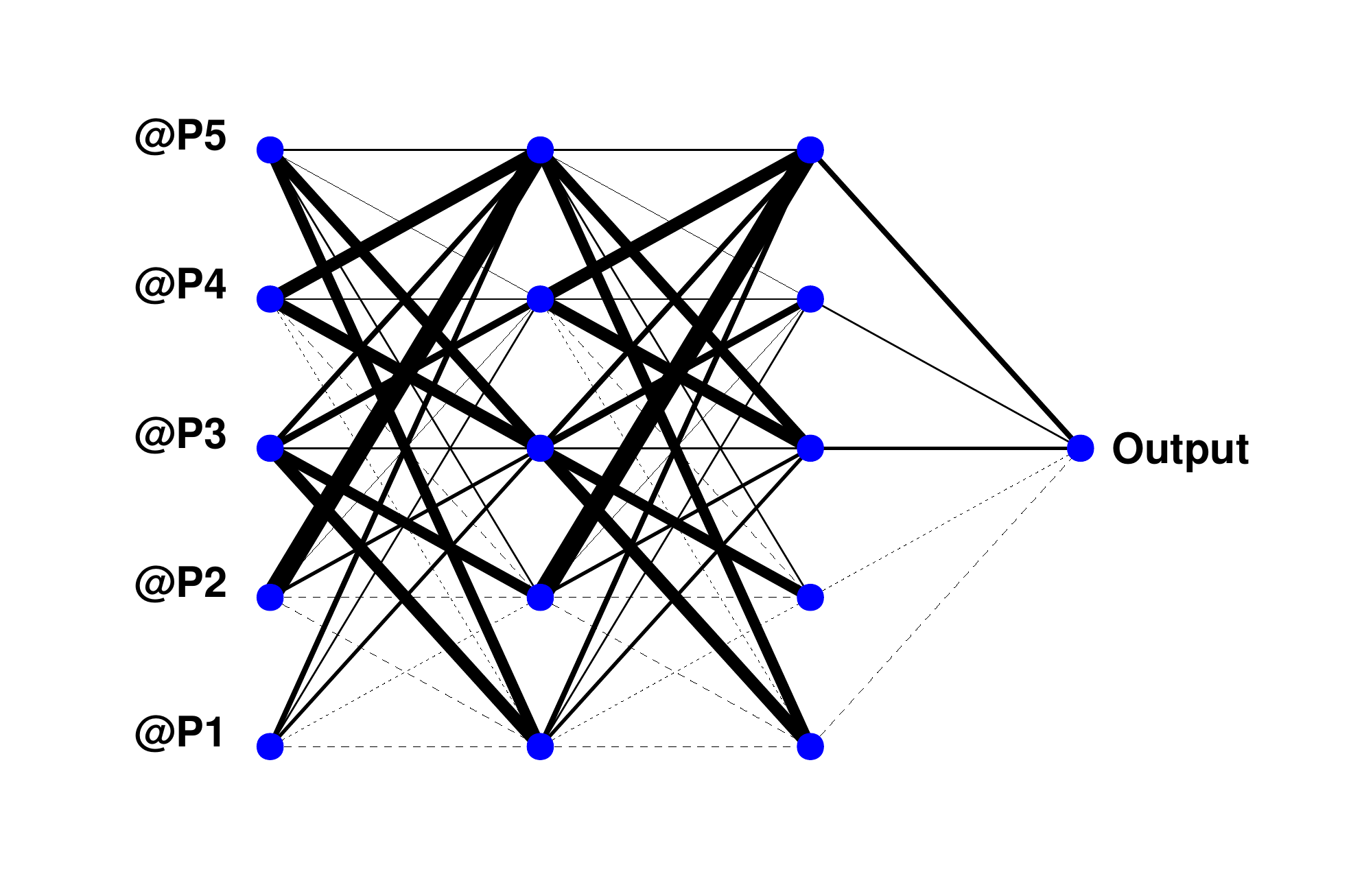}{Architecture of NN}{Fig:estnn}{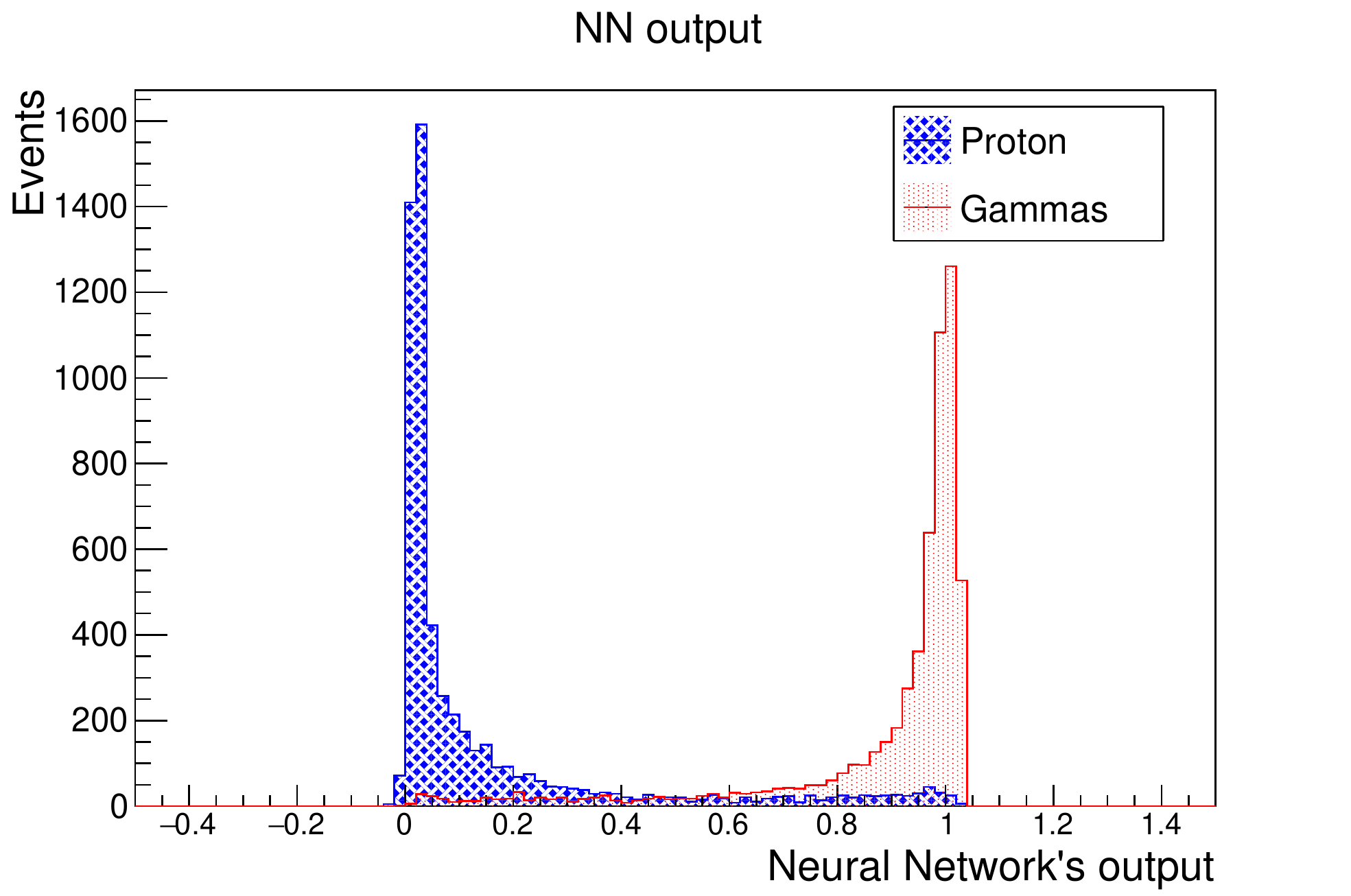}{Outputs of NN}{Fig:outputnn}{In (a) is shown the architecture of NN with 5 neurons as inputs, two hidden layers with 5 neurons and one neuron as output. The width of each connection line between neurons is proportional to the weight of the NN. In (b) is shown the outputs of the NN for gammas and hadrons in the learning stage. The majority of gamma events have an output close to one, and protons are close to 0}{Fig:enn}
		\paragraph*{} In Figure~\ref{Fig:outputnn} is shown the histogram of the output for the NN. The majority of the events produced by GR are close to value 1 and CR to 0. Finding the optimal cut in this variable will allow us to separate between different types of primary particles. This threshold value is defined as $\theta_{NN}$.
		\paragraph*{} The Q factor is defined as $\epsilon _{gamma}~/~\sqrt {\epsilon _{hadron}}$ where $\epsilon _{gamma}$ is the fraction of gamma events that are classified correctly, also called gamma~efficiency, and $\epsilon _{hadron}$ is the hadron events that are classified as gamma events, also called hadron~efficiency. The Q value estimates the factor by which the significance will be increased by the classification. Figure~\ref{Fig:fqentre} shows the Q factor and the $\theta_{NN}$ value, where it can be seen that the highest value of Q corresponds to a value around $\theta_{NN}= 0.98$. The receiver operating characteristic (ROC) curve is useful for comparing classifiers and visualizing their performance \cite{fawcett2006roc}. From the ROC curve we can see that by using $\theta_{NN}=0.96$ we increase the gamma~efficiency, even if we have a bit lower Q Factor with this cut (see Table~\ref{Tab:optiepsi}).
		\doublefig{8}{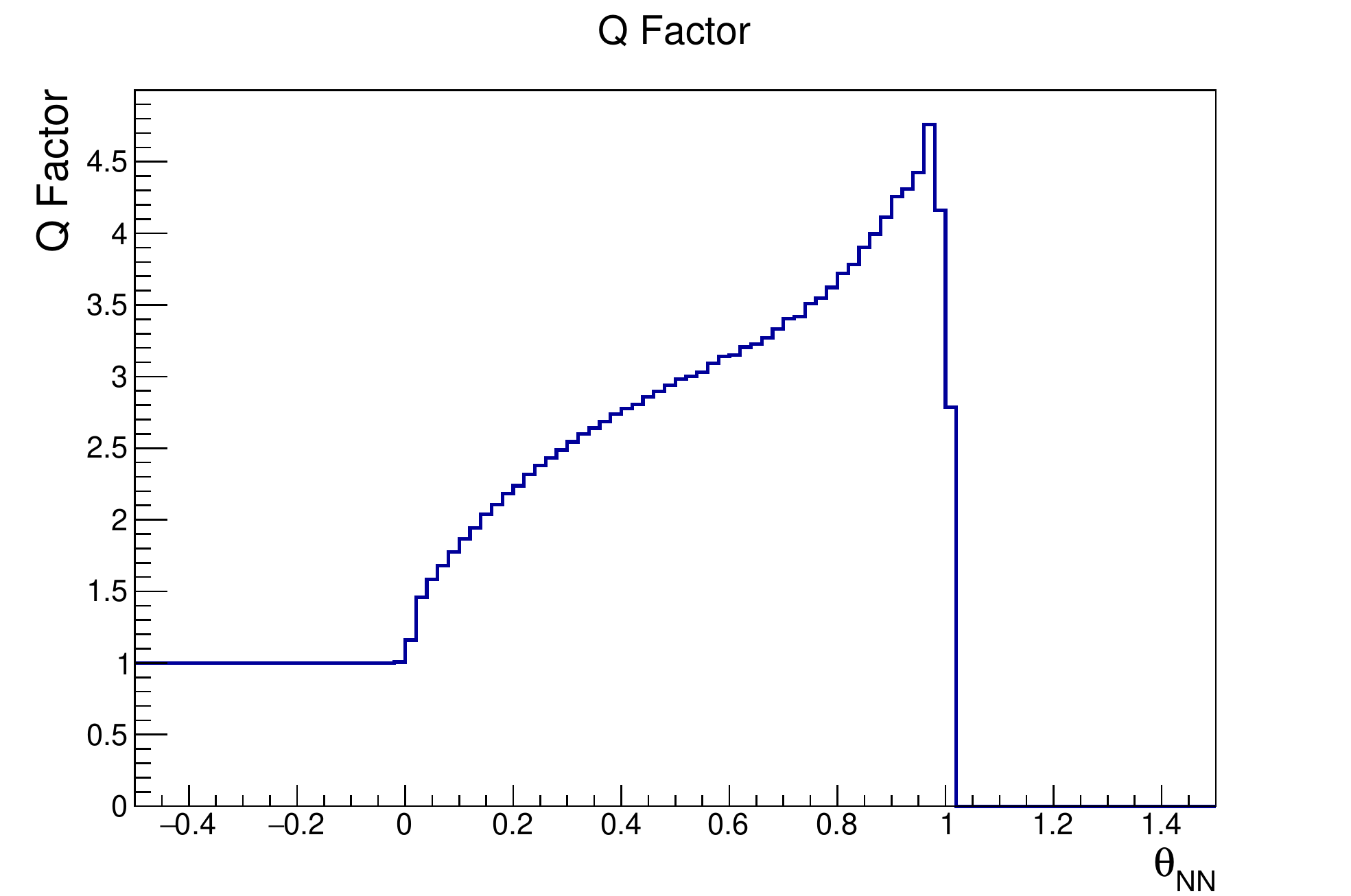}{Q Factor}{Fig:fqentre}{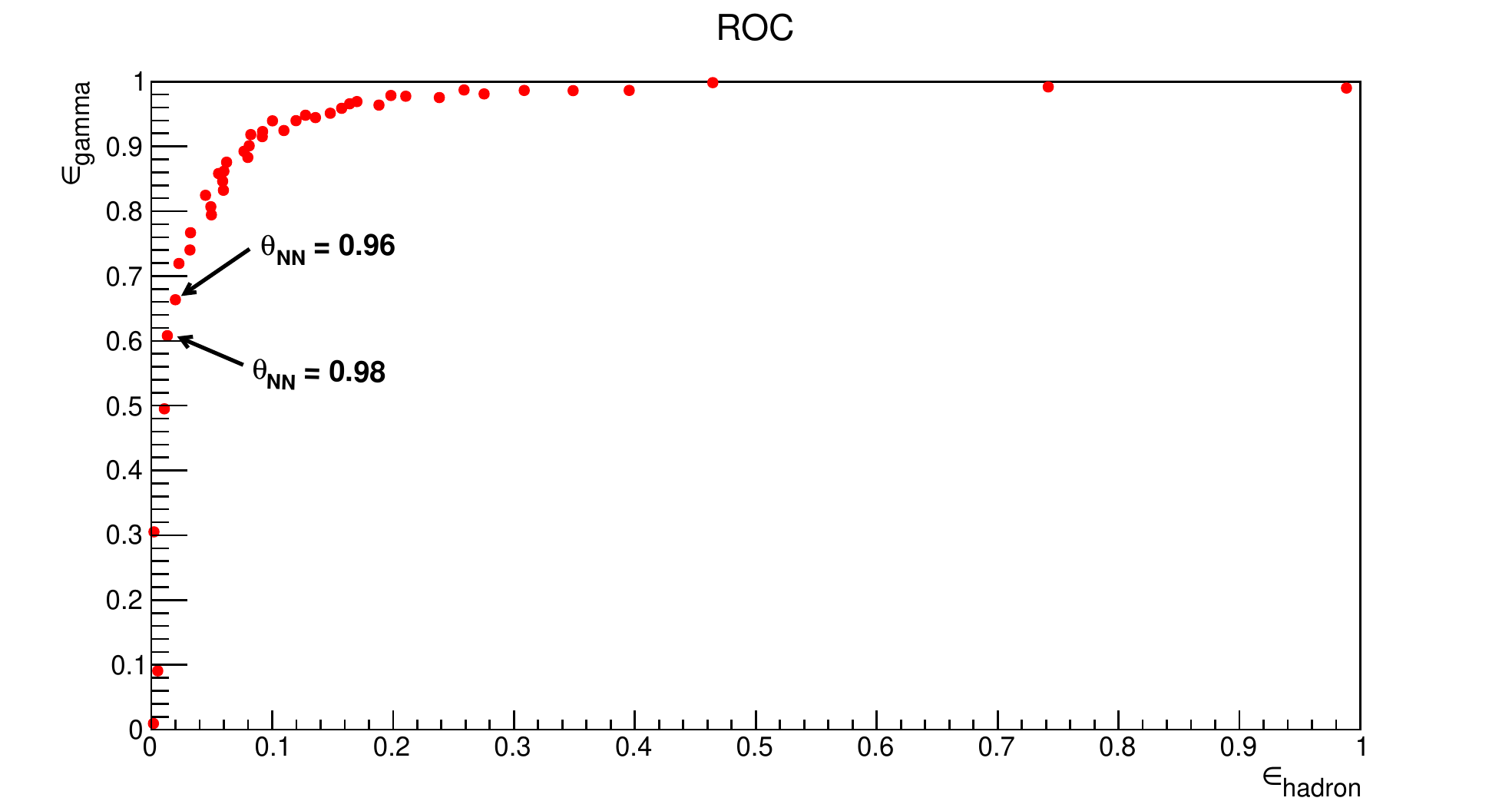}{ROC curves}{Fig:rocentre}{In (a) is shown the Q Factor of NN's outputs. The largest Q factor is at 4.76 when the output threshold is around 0.98. In (b) are shown the ROC for the NN. The $\theta _{NN}$ corresponding to the $\epsilon _{gamma}$ between 0.6 and 0.7 could been used, at a loss of some Q value}{Fig:obtepsi}
	 	\esqtab{Values for gamma and hadron efficiency close to the maximum value of Q factor. Here, for completeness, we include the bin -1 from Table~\protect\ref{Tab:hawc300}, even thought the bin is not used in the compactness analysis}{Tab:optiepsi}{|c|c|c|c|}{
			\hline
			$\epsilon _{NN}$ & $\epsilon _{gamma}$ & $\epsilon _{hadron}$ & Q Factor\\
			\hline
			0.94 &  0.713 &  0.028 &  4.309\\
			\hline
			0.96 &  0.666 &  0.024 &  4.424\\
			\hline
			0.98 &  0.604 &  0.019 &  4.761\\
			\hline
			1.00 &  0.495 &  0.011 &  4.160\\
			\hline
			1.02 &  0.306 &  0.005 &  2.787\\
			\hline
		}
		\subsection{Choice of characteristic inputs}
			\paragraph*{}The main idea is to use the morphological differences of the charge distribution in the PMTs for the two type of primary particles. In event produced by gammas the PMTs close to the core of the shower have the biggest signals and the charge distribution is characterized by a compact and smooth profile. But in the case of hadrons, PMTs with high charge can be far away from the core and the charge distribution is not compact.
			\begin{itemize}
				\item The first feature we include is the number of PMTs with at least one photoelectron (PE) because it is directly related to the energy of the primary. We need our NN to distinguish independently of the energy of the CR or GR. This replaces the nHit binning used with the compactness cut (P1=nHit).
				\item $DisMax$ (P2) that is the largest distance between any of the pair of tubes passing the next selection:
first all the PMTs in the event are sorted by their PEs detected and we summed this value for each PMT from higher to lower until the sum is less that $(SumPE-MaxPE)k(nHit)$, where $MaxPE$ is the number of PEs in any PMT in the event, and "k" is a factor that depends linearly of nHit, the PMTs involved in that sum are the selected ones. This input involves the distance of the PMTs with biggest charge detected and its distance because we suppose that for gammas all the PMTs with high PE are neighbors and this $DisMax$ should be small.
				\item $P3$ this feature is associated with the integral of the radial density where the hadron shower dominates gamma shower \cite{Art:Grabski2011} defined as:
					\begin{center}
						$Log_{10} (\frac{nHit}{\sum_{n}PE_i * R_{PE_i}})~~~~~~~~~~where~R_{PE_i}~>~30~m$
					\end{center}
                    Here $PE_i$ is the charge in the PMT$_i$,  $R_{PE_i}$ is the distance in meters between the PMT$_i$ and position of the reconstructed shower center (core).
				\item $P4$ is defined as $CxPE_{30}/MaxPE$, where $CxPE_{30}$ is the maximum charge outside a exclusion radius of 30 m in the event. For protons one expects to often see charge localized high charge deposition far from the core, so P4 can approach 1 for protons. On the other hand, gammas usually have a value near 0 because most of their charge is deposited near the core.
				\item $P5$ is related to the difference between the maximum charge outside and inside the exclusion radius weighted with the distance to the core.
				    \begin{center}
						$P5=Log_{10}(|CxPE_{30}*R_{CxPE_{30}}-PE_{maxint}*R_{PE_{maxint}}|)$
						$where~R_{CxPE_{30}}~>~30~m~and~R_{PE_{maxint}}~<~30~m$
					\end{center}
			\end{itemize}
		\subsection{Training data set}
			\paragraph*{} The simulated events were generated by using CORSIKA program in the energy range [0.005,100] TeV with a flat spectrum and zenith angle $[0,75]^o$. The performance and response of the array were computed using the HAWC official software.	
			\paragraph*{}For the training stage, the network need two data sets, one for gamma and other for hadron. We defined a target value of 1 for gamma ray event and 0 for hadron event. In this work we only use protons as hadrons because protons constitute nearly 99\% of the CR.
The conditions for selecting training the events for each set are:
				\begin{itemize}
					\item The event is well reconstructed.
					\item The difference between the core reconstruction and simulation does exceed 5 m. 
					\item The core falls inside the HAWC array  
					\item Event with nHit between 30 and 1200.  
				\end{itemize}
	\section{Testing stage}
		\subsection{Simulation}
			\paragraph*{}In this stage we use the same criteria described above for selecting the events for the training data set which consists of new simulated events independent of the training set. For comparing the two methods we use events with $55 < nHit < 1200$, that correspond to bin 0 up to bin 9, i.e. we are not using the bin $-1$. 
In this comparison we will simply weight all events equally, without the optimal weighting for events in each bin used in \cite{pacoicrc2015} the Crab analysis. However, we do apply the compactness cuts of Table~\ref{Tab:hawc300} for each nHit bin to compare performance of the NN and compactness. 
The bin called "total" is computed using all events from bin 0 to bin 9. The results are shown in Figure~\ref{Fig:fq96} where we can see that for the Q Factor the NN has a better result than using the compactness method.
			\onefig{8}{8}{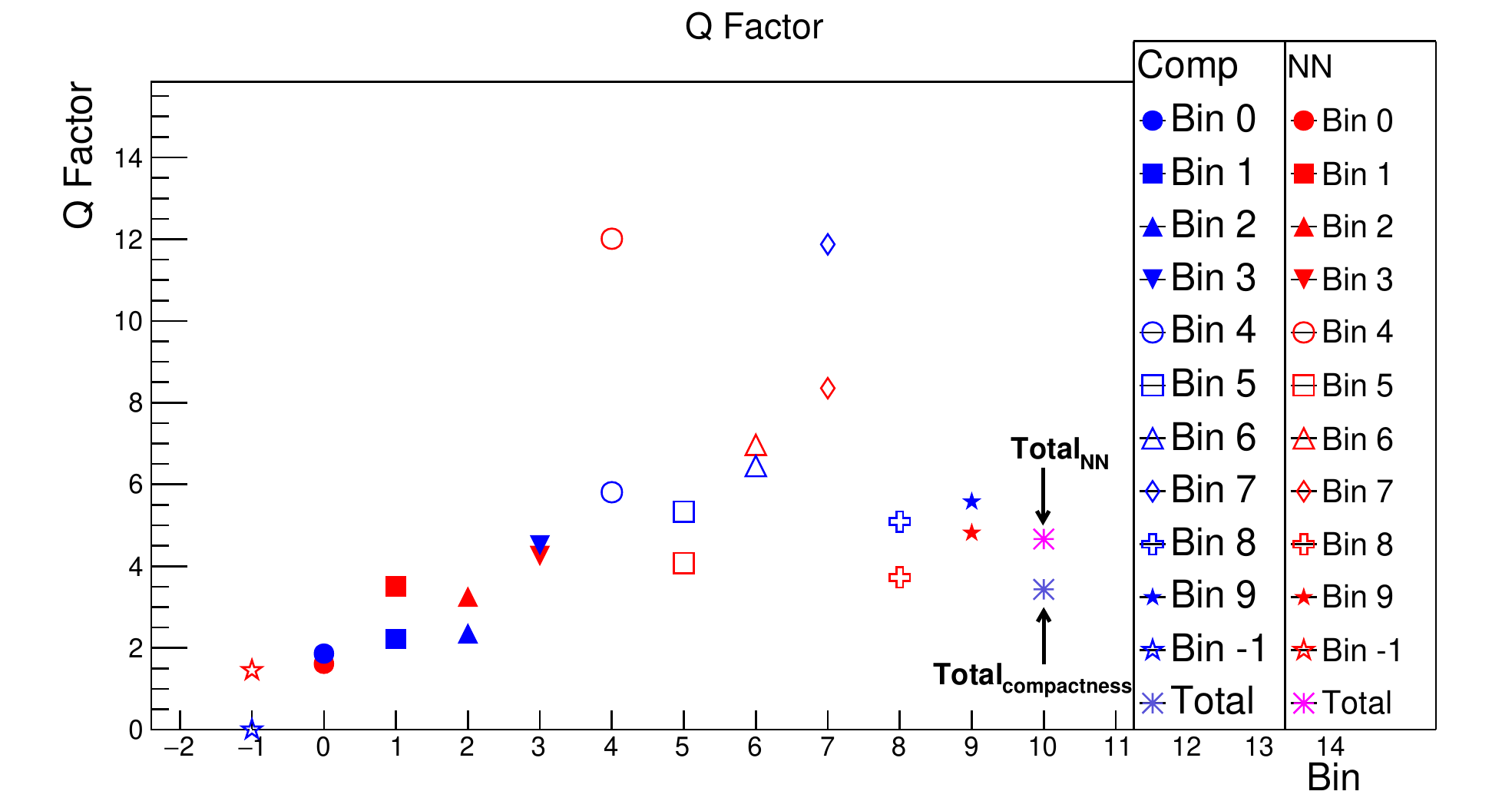}{The Q factor is calculated for each bin and the total (bin 0 to 9) with $\theta _{NN}=0.96$. This shows that for the Q factor in some bins, the NN is better than compactness but for others does not. Using the total bin we got $60\%$ and $53\%$ in gamma efficiency for NN and compactness respectively}{Fig:fq96}
			\paragraph*{}The total value of Q Factor , gamma efficiency and hadron efficiency of each separation methods (compactness and NN) is shown in the Table~\ref{Tab:parasimula}. The NN improves on the compactness method. The gamma efficiency increased by $13\%$ and the hadron efficiency decreased $30\%$, so the Q factor increased by $35\%$.
			\esqtab{Difference between methods with simulation}{Tab:parasimula}{|c|c|c|c|}{
				\hline
				Parameter & NN & compactness & Increase (\%)\\
				\hline
				Q Factor  & 4.663 & 3.432 &  35.889\\
				\hline
				gamma efficiency      & 0.606 & 0.536 &  13.129\\
					\hline
				hadron efficiency       & 0.017 & 0.024 & -30.693\\
					\hline
			}
		\subsection{Data}
			\paragraph*{}Another way to compare the different performance of the compactness and the NN is using HAWC data. We chose a set of well reconstructed events within $\pm 6 ^{\circ}$ of the Crab Nebula. We have two methods (NKG and Gauss) for reconstructing the core position, but only Gauss was used in training the NN. A well-behaved event should have a similar core position for either method. In the case of using compactness we use a very simple analysis \cite{john2014} and apply a cut of $\theta _c$ that varies from 10 to 18  but is applied to all nHit bins. For technical reasons we were not able to apply the bin-dependent cuts of Table~\ref{Tab:hawc300} to the Crab data, so this constitutes a preliminary comparison of NN and compactness on the Crab data. The results are shown in Table~\ref{Tab:sigmacompa}. In the case of NN method, the maps are obtained by varying  $\theta _{NN}$ from $0.92$ to $1.0$ (see Table~\ref{Tab:sigmann}).
	%tabla compactness
			\esqtab{Significance using the compactness variable with a single cut value for all bins}{Tab:sigmacompa}{|c|c|c|}{
				\hline
				$\theta _c$ & NKG & Gauss\\
				\hline
				10.0 &  3.4706 &  4.4649\\
				\hline
				12.0 &  4.3142 &  4.4703\\
				\hline
				14.0 &  5.2777 &  4.6895\\
				\hline
				16.0 &  3.9327 &  4.3406\\
				\hline
				18.0 &  4.3170 &  4.3613\\
				\hline
			}		
	%tabla nn
			\esqtab{Significance using NN Vs NN threshold}{Tab:sigmann}{|c|c|c|}{
				\hline
				$\theta _{NN}$ & NKG & Gauss\\
				\hline
				0.92 &  5.8842 & 4.9889\\
				\hline
				0.94 &  5.7042 & 5.4144\\
				\hline
				0.96 &  5.9217 & 5.5096\\
				\hline
				0.98 &  3.7534 & 4.6703\\
				\hline
				1.00 &  4.0977 & 3.1792\\
				\hline
			}
			\paragraph*{}The highest values of significance from Tables \ref{Tab:sigmacompa} and \ref{Tab:sigmann} are placed in Table~\ref{Tab:sigmadosme} and the increase with respect to the compactness method is computed. The results show that the NN  is better than compactness in this preliminary comparison, consistent with expectations from MC. With the NKG method, the increase is $12~\%$, and with Gauss method is $17~\%$, not surprising since the NN learnt with events whose core reconstruction was done with Gauss method.
			\esqtab{Difference between methods with data}{Tab:sigmadosme}{|c|c|c|}{
				\hline
				Method & NKG & Gauss\\
				\hline
				compactness & 5.2777 & 4.6895\\
				\hline
				NN & 5.9217 & 5.5096\\
				\hline
				Increase (\%)& 12.202 & 17.488 \\
				\hline
			}
	\section{Conclusion}
		\paragraph*{}In this work, we propose a new method for gamma/hadron separation that used a Multilayer Perceptron fed with 5 characteristics. The NN's output is continuous and has a value targeting 1 for gamma events and 0 for hadron events. In the analysis, we found an optimal cut value for the NN output $\theta_{NN}= 0.96$. With this value the NN has  better performance than compactness. The  Q Factor increases approximately $36 \%$, because the gamma efficiency increased about $13 \%$ and a decrease of $30 \%$ in hadron efficiency.
		\paragraph*{}In the case of Crab data we also obtained a better significance using NN instead of a simplified version of compactness where the compactness cut was constrained to be the same for all nHit bins. In future work we will compare with the full compactness implementation.

	\section*{Acknowledgments}
\footnotesize{
We acknowledge the support from: the US National Science Foundation (NSF);
the US Department of Energy Office of High-Energy Physics;
the Laboratory Directed Research and Development (LDRD) program of
Los Alamos National Laboratory; Consejo Nacional de Ciencia y Tecnolog\'{\i}a (CONACyT),
Mexico (grants 239762,260378, 55155, 105666, 122331, 132197, 167281, 167733);
Red de F\'{\i}sica de Altas Energ\'{\i}as, Mexico;
DGAPA-UNAM (grants IG100414-3, IN108713,  IN121309, IN115409, IN111315);
VIEP-BUAP (grant 161-EXC-2011);
the University of Wisconsin Alumni Research Foundation;
the Institute of Geophysics, Planetary Physics, and Signatures at Los Alamos National Laboratory;
the Luc Binette Foundation UNAM Postdoctoral Fellowship program.
}
	\bibliographystyle{JHEP}
	\bibliography{proceeding}

\end{document}